\RequirePackage[left]{lineno}
\documentclass[twocolumn,superscriptaddress,showpacs,preprintnumbers,amsmath
,amssymb,aps,prd,nofootinbib]{revtex4-1}
\usepackage{graphicx}
\usepackage{slashbox}
\usepackage{bm}
\usepackage{color}
\usepackage{longtable}
\usepackage{hyperref}
\graphicspath{{Figures/}}

\usepackage{float}
\usepackage{amsmath}
\usepackage{multirow}
\usepackage{ifthen}
\usepackage{blindtext}
\usepackage{svn-multi}
\usepackage[normalem]{ulem}
\usepackage{hyperref}
\usepackage{etoolbox}
\usepackage{fancyhdr}
\usepackage{import}
 \usepackage{etoolbox}

\newtoggle{fullauthorlist}
\toggletrue{fullauthorlist}

\newtoggle{endauthorlist}
\toggletrue{endauthorlist}

\newcommand{\dd}{\ensuremath{\mathrm{d}}}

\usepackage{braket}

\usepackage[symbol]{footmisc}
\renewcommand{\thefootnote}{\fnsymbol{footnote}}

\begin{document}

\title{Search for intermediate mass black hole binaries\\
 in the first observing run of Advanced LIGO}
 
\iftoggle{endauthorlist}{
  %
  %
  \let\mymaketitle\maketitle
  \let\myauthor\author
  \let\myaffiliation\affiliation
  \author{B. P. Abbott et al. \footnote[1]{The full author list can be found at the end of the article.}\\
  (LIGO Scientific Collaboration and Virgo Collaboration)}
}{
  %
  %
  \iftoggle{fullauthorlist}{
  \input{O1_IMBHB_LVC_Feb2017-prd}
  }{
    \author{The LIGO Scientific Collaboration and the Virgo Collaboration}
  }
}
\renewcommand*{\thefootnote}{\arabic{footnote}}
\setcounter{footnote}{0}

\date{\today}

\begin{abstract} 
During their first observational run, the two Advanced LIGO detectors attained an unprecedented sensitivity,
resulting in the first direct detections of gravitational-wave signals 
produced by stellar-mass binary black hole systems.  This paper reports on an
all-sky search for gravitational waves (GWs) from merging intermediate mass black hole binaries (IMBHBs). 
The combined results from two
independent search techniques were used in this study: the first employs a matched-filter algorithm that
uses a bank of filters covering the GW signal parameter space, while the second is a generic search for GW
transients (bursts).
No GWs from IMBHBs were detected; therefore, we constrain the rate of several classes of IMBHB mergers.
The most stringent limit is obtained for black holes of individual mass $100\,M_\odot$, 
with spins aligned with the binary orbital angular momentum. For such systems, the merger rate
is constrained to be less than $0.93~\mathrm{Gpc^{-3}\,yr}^{-1}$ in comoving units at the $90\%$ confidence
level, an improvement of nearly 2 orders of magnitude over previous upper limits.
\end{abstract}


\maketitle
\section{Introduction}

The first observing run (O1) of the Advanced Laser Interferometer Gravitational-Wave Observatory (LIGO) detectors~\cite{TheLIGOScientific:2014jea} took place from September 12, 2015 to January 19, 2016. During this period, there were
a total of $51.5~\mathrm{days}$ of coincident analysis time between the two detectors, located in 
Hanford, Washington (H1), and Livingston, Louisiana (L1).  This resulted in the detection of gravitational-wave (GW) signals
from the coalescence of two binary black hole (BBH) systems with high statistical significance, GW150914~\cite{Abbott:2016blz} and 
GW151226~\cite{Abbott:2016nmj}, and a third lower-significance candidate, LVT151012~\cite{TheLIGOScientific:2016qqj}, which is also
likely to be a BBH coalescence~\cite{TheLIGOScientific:2016pea}.

In all three cases, the estimated premerger individual source-frame masses, $(36.2^{+5.2}_{-3.8},29.1^{+3.7}_{-4.4})M_{\odot}$, 
$(14.2^{+8.3}_{-3.7},7.5^{+2.3}_{-2.3})M_{\odot}$, and $(23^{+18}_{-6},13^{+4}_{-5})M_{\odot}$, 
respectively~\cite{TheLIGOScientific:2016wfe,TheLIGOScientific:2016pea}, are consistent
with stellar evolutionary scenarios~\cite{TheLIGOScientific:2016htt}.\footnote{Since GWs undergo a cosmological redshift between source and detector, we relate the observed detector-frame mass $m_\mathrm{det}$ and the physical source-frame mass $m_\mathrm{source}$ via $m_\mathrm{det}=(1+z)m_\mathrm{source}$, where $z$ is the redshift of the source assuming standard cosmology~\cite{Planck:2016}.}  
These systems were observed at relatively low redshifts,
$z = 0.09^{+0.03}_{-0.04}$, $0.09^{+0.03}_{-0.04}$, and $0.20^{+0.09}_{-0.09}$, respectively.  If relatively heavy black hole remnants, similar to those already observed by Advanced LIGO, exist within dense
globular clusters (GCs), further hierarchical merging of these objects could be a natural formation mechanism
for intermediate mass black holes (IMBHs)~\cite{Miller:2003sc}. IMBHs are normally 
defined as black holes with masses in the range $10^2\leq M_{\bullet}/M_{\odot}\leq10^5$; in this paper, we consider any BBH with a total
mass above $10^2\,M_{\odot}$ and mass ratio of $0.1\leq q\leq 1$ to be an IMBH binary (IMBHB).

It is possible that there will be numerous BBH detections in the next few years of GW
astronomy~\cite{Abbott:2016nhf,TheLIGOScientific:2016pea,Abbott:2016ymx}. In the near future, we should be able 
to place stringent astrophysical constraints on the formation and evolution of stellar-mass black holes.  In addition to surveying stellar-mass black holes, we will also be able to investigate the astrophysics of IMBHs.  

\begin{figure}[thb]
  \includegraphics[scale=0.42,trim ={10 110 30 130}, clip]{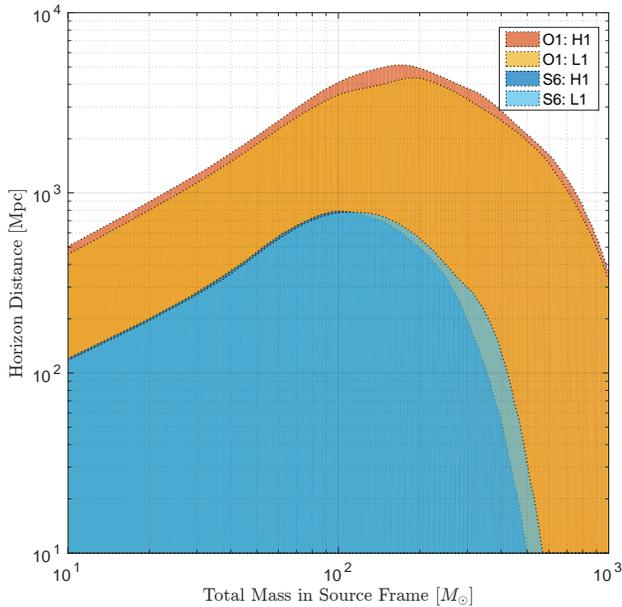}
  \caption{
  Horizon distance for equal-mass, nonspinning binary black hole systems with a single-detector SNR threshold of $8$ in the first observing run (O1) of the Advanced LIGO detectors~\cite{TheLIGOScientific:2016agk}.  Comparison
  curves are also given for the previous sixth science run (S6) of the initial LIGO detectors~\cite{Abbott:2007kv}.
  }
  \label{f:O1horizon}
\end{figure}

If they are found to exist, IMBHB mergers would be the LIGO--Virgo sources that emit the most gravitational-wave energy.  Given an estimate of the power spectral density of a detector~\cite{TheLIGOScientific:2016agk}, and assuming a matched-filter single-detector signal-to-noise ratio (SNR) threshold of $8$, in Fig.~\ref{f:O1horizon}, we plot the horizon distance (the distance to which we can detect an optimally located and oriented source) as a function of source-frame total mass.  As Fig.~\ref{f:O1horizon} displays, the O1 sensitivity for IMBHBs constitutes a 
factor of $\approx6$ improvement in peak horizon distance ($\approx200$ in search volume) as compared to the sensitivity achieved between 2009--2010, during the sixth and final science run (S6) of initial LIGO~\cite{Aasi:2014iwa}.  However, the matched-filter SNR is only an optimal detection statistic in stationary, Gaussian noise.  Since LIGO data are known to contain non-stationary noise~\cite{TheLIGOScientific:2016zmo}, this figure is useful primarily as an approximate upper bound on the reach of a modeled search for IMBHBs.

\begin{table}[!htp]
  \caption{Results of our analysis for IMBHB systems with (source-frame) component masses $m_{1,2}$ and spins $\chi_{1,2}$ parallel to the orbital angular momentum. For each set of parameters, we report the $90\%$ confidence combined upper limit on the rate density $R_{90\%}$ and the combined- and single-pipeline sensitive distance $D_{\langle VT\rangle}$. Uncertainty in the detectors' amplitude calibration introduces a $\approx18\%$ uncertainty in the rates and a $\approx6\%$ uncertainty in the sensitive distance.}
\begin{center}
\begin{ruledtabular}
  \begin{tabular}{ c c c  c c c}
      $m_{1}$ & $m_{2}$ & $\chi_{1,2}$ & \multicolumn{2}{c}{$R_{90\%}$} & $D_{\langle VT\rangle} \mathrm{{}^{GstLAL}_{cWB}}$ \\           
    $[M_{\odot}]$ & $[M_{\odot}]$ & & \multicolumn{1}{c}{$[\mathrm{Gpc^{-3}\,yr}^{-1} ]$} &  \multicolumn{1}{c}{$[\mathrm{GC^{-1}\,Gyr}^{-1} ]$} & $[\mathrm{Gpc}]$ \\ \hline

$100$ & $100$ & $0.8$ & $0.93$  & $0.3$ & $1.6~^{1.7}_{1.3}$ 
\\
$100$ & $100$ & $0$ & $2.0$  & $0.7$ & $1.3~^{1.3}_{1.0}$ 
\\
$100$ & $100$ & $-0.8$ & $3.5$  & $1$ & $1.1~^{1.1}_{0.89}$ 
\\
$100$ & $20$ & $0$ & $13$  & $4$ & $0.68~^{0.69}_{0.46}$ 
\\
$100$ & $50$ & $0$ & $3.3$  & $1$ & $1.1~^{1.1}_{0.78}$ 
\\
$200$ & $50$ & $0$ & $9.8$  & $3$ & $0.75~^{0.76}_{0.66}$ 
\\
$200$ & $100$ & $0$ & $4.6$  & $2$ & $0.97~^{0.98}_{0.84}$ 
\\
$200$ & $200$ & $0$ & $5.0$  & $2$ & $0.94~^{0.95}_{0.78}$ 
\\
$300$ & $50$ & $0$ & $45$  & $20$ & $0.45~^{0.46}_{0.37}$ 
\\
$300$ & $100$ & $0$ & $16$  & $5$ & $0.63~^{0.64}_{0.52}$ 
\\
$300$ & $200$ & $0$ & $12$  & $4$ & $0.69~^{0.70}_{0.58}$ 
\\
$300$ & $300$ & $0$ & $20$  & $7$ & $0.59~^{0.60}_{0.45}$ 
\\

  \end{tabular}
\end{ruledtabular}
\end{center}
  \label{tab1}
\end{table}

In this paper, we report on the search for IMBHBs during O1.
In previous IMBHB searches using LIGO--Virgo data taken in 2005--2010~\cite{Aasi:2014iwa, Aasi:2014bqj}, an unmodeled transient search and a modeled matched-filter search using only the ringdown part of the 
waveform were separately employed to set distinct upper limits on the merger rates of IMBHBs.
For this study, two distinct search pipelines were also used: a 
matched-filter search algorithm, GstLAL~\cite{Cannon:2011vi,Privitera:2013xza,Messick:2016aqy}, that uses inspiral--merger--ringdown 
waveform templates~\cite{TheLIGOScientific:2016qqj,TheLIGOScientific:2016pea} which are cross-correlated with the data, and 
an unmodeled transient search algorithm, coherent WaveBurst (cWB)~\cite{TheLIGOScientific:2016uux,Klimenko:2015ypf,Abbott:2016ezn}, which looks
for excess power which is coherent across the network of GW detectors.
Instead of setting distinct upper limits, however, the results
presented in this paper are the combined statistics from both independent search techniques. No IMBHBs were 
detected in this combined search in O1; based on this, we set a $90\%$ confidence level limit on the rate of mergers (see Table~\ref{tab1} below).

The paper is organized as follows: Section~\ref{sec:search} summarizes our search techniques and how they are combined for the current analysis. Section~\ref{sec:sensitivity} describes how upper limits on rates are calculated and includes Table~\ref{tab1} and Fig.~\ref{f:Rates} as main results. Section~\ref{sec:astro} discusses the astrophysical implications
inferred from this analysis, and Section~\ref{sec:conclusion} presents our conclusions. We use the ``TT+lowP+lensing+ext'' parameters from Table~4 of the Planck 2015 results~\cite{Planck:2016} for cosmological calculations.

\section{Search Technique}
\label{sec:search}
For O1, a 
new search was inaugurated, in which
both modeled and unmodeled analyses, specifically tuned to search for IMBHBs, were combined to form a single search. 
The modeled analysis employs a matched filter, which yields the optimal detection 
efficiency for signals of known form in stationary, Gaussian noise~\cite{Wiener:1949} and thus requires a sufficiently accurate signal waveform model for use as a template. 
The unmodeled transient analysis, by contrast, can identify burst-like signals which do not correspond to any currently available waveform model. 
IMBHB signals, as a consequence of their sources' high mass, have relatively few cycles in the LIGO frequency band; therefore, the IMBHB search benefits from the combination of the two complementary analysis techniques.

\subsection{Modeled Analysis}
\label{sec:modeled_analysis}
The GstLAL pipeline, which is a matched-filter search algorithm for GWs from compact binary coalescences~\cite{Cannon:2011vi,Privitera:2013xza,Messick:2016aqy}, was used in its offline mode to analyze the entirety of O1~\cite{TheLIGOScientific:2016qqj,TheLIGOScientific:2016pea}.  
\begin{figure*}[thp]
 \centering
 \begin{minipage}[b]{.45\textwidth}
  \includegraphics[scale=0.47]{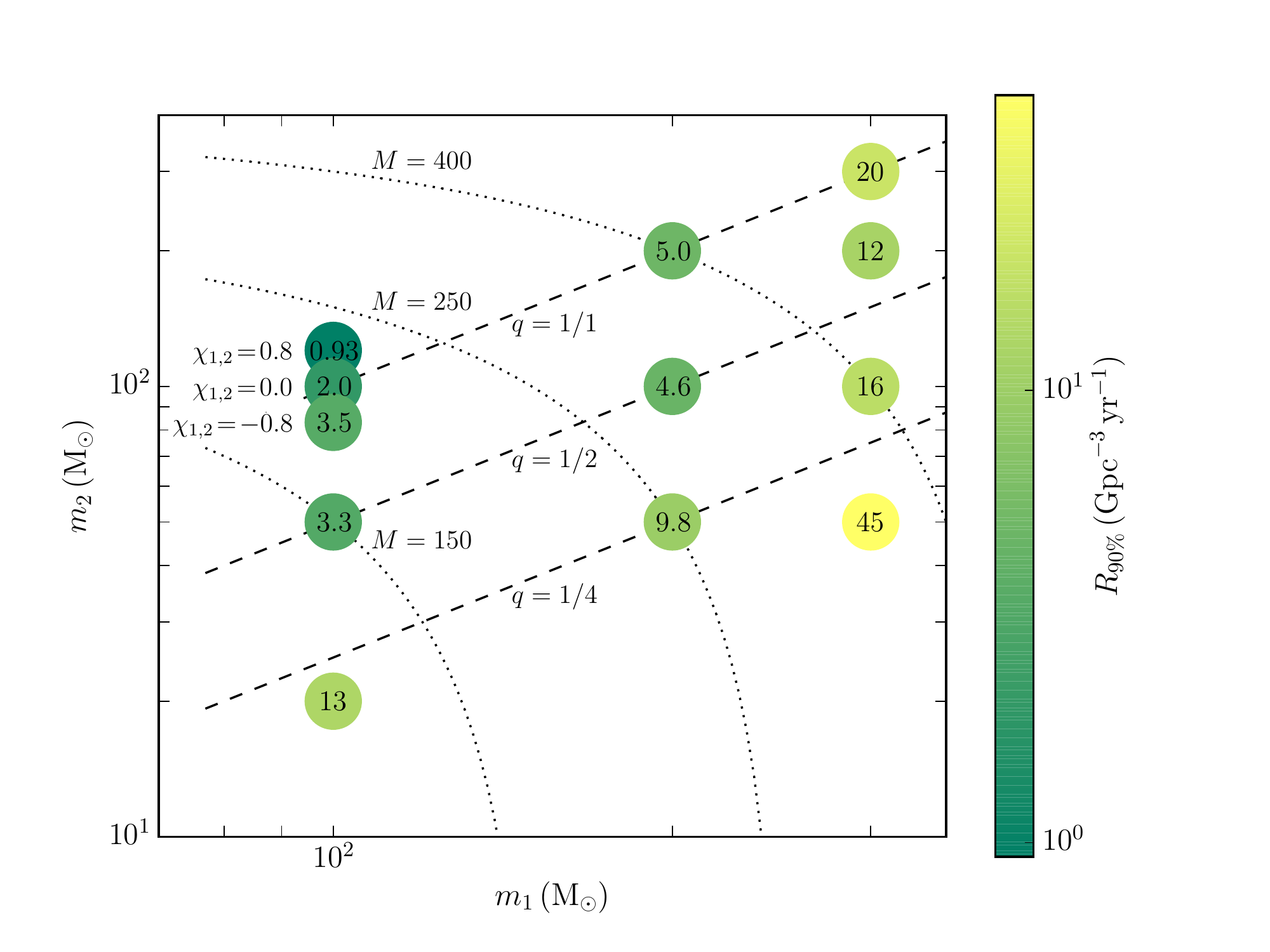}
  Rate upper limit 
  \end{minipage}
 \qquad
 \begin{minipage}[b]{.45\textwidth}
  \includegraphics[scale=0.47]{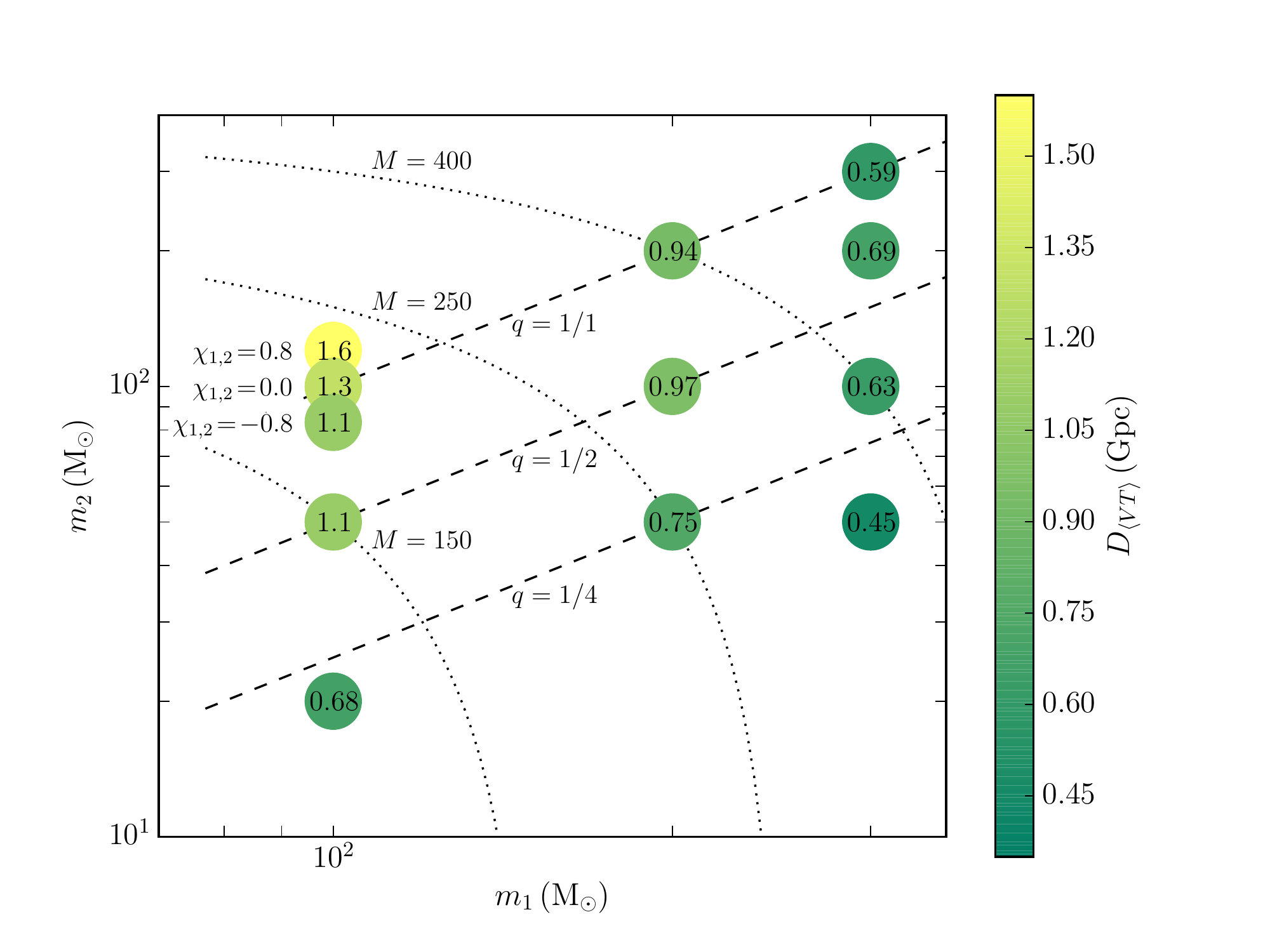}
  Sensitive distance 
 \end{minipage}
   \caption{$90\%$ confidence rate upper limit in $\mathrm{Gpc^{-3}\,yr^{-1}}$ 
   (left panel) and sensitive distance in $\mathrm{Gpc}$ (right panel) achieved by this search for IMBHB mergers in Advanced LIGO's
   first observing run.  Each circle represents a set of simulated IMBHB signals, with circles centered on the component masses $(m_1,m_2)$.  All except two sets have
   nonspinning binary components.  For masses 
   $m_1 = m_2 = 100\,M_\odot$, additional simulations were performed with spins aligned ($\chi_1=\chi_2=0.8$) or
   antialigned ($\chi_1=\chi_2=-0.8$) with the orbital angular momentum; these are shown as displaced circles.
   The straight dashed lines represent contours of constant mass ratio $q=m_2/m_1$; the curved dotted lines are those of constant total mass $M = m_1 + m_2$.
   All reported quantities are calculated in the source frame.}
   \label{f:Rates}
\end{figure*}

The GstLAL IMBHB analysis is based on a discrete bank of GW templates constructed over a total mass between $50\, M_\odot$ and $600\, M_\odot$ in the detector frame, with mass ratios less extreme than $1$:$10$, and with dimensionless spin $\chi_{1,2}$ between $-0.99$ and $0.99$, where positive values are aligned with the orbital angular momentum of the system and negative values are antialigned.  The templates used in this search are a reduced-order model of a double aligned-spin effective-one-body waveform~\cite{SEOBNRv2,Purrer:2015tud}.  As a consequence of the noise characteristics at low frequencies~\cite{TheLIGOScientific:2016agk}, GstLAL began its analysis at a frequency of $15~\mathrm{Hz}$.

In this analysis, the data are filtered through a singular-value decomposition of the template bank, and the matched-filter SNR time series for each template in the bank is reconstructed from the filtered output of the basis templates~\cite{Messick:2016aqy}. Maxima in the SNR, called triggers, are identified, and corresponding values of a signal consistency test, which is a comparison of the SNR time series for the data to the SNR time series expected from a real signal, are computed.  Triggers found in one detector that are not coincident with triggers in another detector are assumed to be non-astrophysical and are used to estimate the probability distribution of noise events in each detector.  Coincident triggers are considered GW candidates and are ranked against each other via a likelihood ratio, which compares the probability that each is a signal to the probability that each is noise~\cite{Messick:2016aqy}. Finally, a coincident trigger is assigned a $p$-value~\cite{Messick:2016aqy}, which is the probability of finding a noise fluctuation with such likelihood ratio or higher under the hypothesis that the data contains no GW signals.\footnote{See Ref.~\cite{Capano:2016uif} for a study of the properties of different methods to estimate the $p$-value in a coincident search for transient GW signals.}

For validation, another independent matched-filter search algorithm, PyCBC~\cite{Canton:2014ena,Usman:2015kfa}, was also run over the same GW parameter space using a spin-aligned frequency-domain phenomenological waveform model~\cite{Husa:2015iqa,Khan:2015jqa} as templates.  
PyCBC uses a different SNR-based ranking statistic~\cite{Canton:2014ena,Usman:2015kfa,Babak:2012zx,TheLIGOScientific:2016qqj}.  These two independent matched-filter algorithms find consistent results over the IMBH parameter space, which increases our confidence in their reliability and robustness.

The three most significant events from the GstLAL matched-filter analysis correspond to GW150914, LVT151012 and GW151226, which have already been reported~\cite{Abbott:2016blz,Abbott:2016nmj,TheLIGOScientific:2016qqj,TheLIGOScientific:2016pea,TheLIGOScientific:2016wfe}. Since parameter-estimation studies have placed these events outside of the IMBH mass range~\cite{TheLIGOScientific:2016wfe,TheLIGOScientific:2016qqj,Abbott:2016izl,TheLIGOScientific:2016pea}, we have removed these triggers from our analysis.  We discuss the production of our overall IMBHB results in Sec.~\ref{sec:combine}.

The bank of waveform templates used by the GstLAL IMBHB analysis notably overlaps with the O1 stellar-mass BBH search~\cite{TheLIGOScientific:2016qqj,TheLIGOScientific:2016pea} between $M=50\, M_\odot$ and $100\, M_\odot$. It was therefore expected that this new analysis would find GW150914 and LVT151012 as two of its most significant events, since the masses of these two signals have posterior support in this range~\cite{TheLIGOScientific:2016qqj,TheLIGOScientific:2016wfe,TheLIGOScientific:2016pea}. Additionally, GW151226 being the third most significant event in this analysis demonstrates the robustness of modeled analyses to identify signals even outside of their covered parameter spaces.

This is the first modeled analysis that includes the inspiral, merger and ringdown portions of the compact binary coalescence waveform to extend above $M=100\, M_\odot$ and into the IMBHB parameter space.  Even though IMBHB mergers potentially have large values of SNR, detecting them with this analysis can be difficult.  Signal consistency checks are often inefficient at distinguishing true signals from background events.  This problem is caused primarily by the short duration of signals produced by high-mass systems, especially those with antialigned spin configurations.  Continuing to pursue improvements in IMBHB search methods will undoubtedly improve the sensitivity of the analysis.

\subsection{Unmodeled Analysis}
\label{sec:unmodeled_analysis}
The unmodeled analysis was conducted with cWB, the data-analysis algorithm used for previous LIGO--Virgo unmodeled IMBHB searches~\cite{Virgo:2012aa,Aasi:2014iwa}. More recently, this algorithm has been used extensively on O1 data~\cite{Abbott:2016ezn}. 

cWB performs a coherent analysis on data from multiple detectors~\cite{Klimenko:2015ypf}; for the O1 analysis, just the H1 and L1 detectors were available. After decomposing the data into a time--frequency representation, the algorithm identifies coherent triggers
from regions in the time--frequency domain with excess power relative to the noise level. GW candidate events are subsequently reconstructed in the framework of a constrained maximum-likelihood analysis. 

As this reconstruction of signal is agnostic to the waveform modeling of the specific astrophysical source, this algorithm can be used in a variety of searches, including
eccentric BBH mergers~\cite{Tiwari:2015gal}. 
Past simulation studies have shown that the cWB unmodeled analysis is sensitive to BBH mergers over large regions of the binary parameter space accessible with initial GW detectors~\cite{PhysRevD.90.022001};
analogous conclusions were reached for the case of advanced detectors~\cite{Mazzolo:2014kta}. 

For this analysis, we applied a further, weak constraint to favor the reconstruction of chirality-polarized waveforms~\cite{Klimenko:2015ypf}. Moreover, with respect to the generic burst search reported in \cite{Abbott:2016ezn}, frequency-dependent post-production selection cuts were tuned in order to minimize the impact of such cuts on IMBHB mergers: the low-frequency part of the spectrum of GW data is often polluted by various environmental and instrumental noises that effectively mimic the expected waveforms for massive binary mergers. 
The cWB analysis began at a frequency of $16~\mathrm{Hz}$. 

The significance of any GW candidate event is estimated by comparing it with the noise background distribution in order to calculate its $p$-value. The background set was empirically produced by
analyzing $\simeq9000$ independent time-shifted O1 data sets.\footnote{Since the noise sources are uncorrelated between H1 and L1, introducing relative time delays larger than the GW travel time  ($\lesssim 10~\mathrm{ms}$) is an effective way to generate 
an empirical noise distribution.} 
Approximately $1100~\mathrm{yr}$ of effective background livetime was accumulated with this procedure.
Additional time lags would have been analyzed had loud IMBHB candidates been identified and a more precise estimate of the background tails been required. 
The only GW signal found in the O1 data by cWB was GW150914, which is louder than all background events. 
Similarly to the aforementioned matched-filter searches, GW150914 was then removed from the unmodeled analysis.

\subsection{Combining Analyses}
\label{sec:combine}

After running on the data collected by the detectors, each search algorithm produces a trigger list with times and associated $p$-values $\mathcal{P}$.
We combine the two lists together to form a single list of triggers ranked by their $p$-value.
To avoid double counting of events, we remove triggers within $100~\mathrm{ms}$ of a more significant trigger found by the other search algorithm.
To account for the use of two search algorithms, we apply a trials factor of $2$ to produce the final $p$-value of our search,
\begin{equation}
\overline{\mathcal{P}} = 1 - (1 - \mathcal{P})^2 .
\end{equation}
This assumes that the triggers produced by the two algorithms are independent; a correlation in the two lists of triggers from the pipelines would reduce the effective trials factor, making $2$ a conservative choice. Of the top $150$ triggers output by the two pipelines, only GW150914 is common between the lists, indicating that the noise triggers are independent here. We consider $\overline{\mathcal{P}}$ as the ranking statistic for the combined search algorithm. Excluding GW150914, LVT151012 and GW151226, the most significant trigger has $\overline{\mathcal{P}} \simeq 0.26$, well below the significance needed to be considered as a detection candidate.

\section{Upper limits on rates}
\label{sec:sensitivity}

Since no IMBHB coalescences were detected during O1, 
we can calculate upper limits on the astrophysical rate (density) of such events.  
With the loudest-event method~\cite{Biswas:2007ni}, if the most significant IMBHB trigger is consistent with noise, 
the $90\%$ confidence upper limit is given by
\begin{equation}
  R_{90\%} = - \frac{\ln(0.1)}{\langle VT\rangle} = \frac{2.303}{\langle VT\rangle},
\end{equation}
where $\langle VT\rangle$ is the averaged spacetime volume to which our 
search is sensitive at the loudest-event threshold. We compute $\langle VT\rangle$ by injecting a large number of simulated waveforms into the O1 data, then analyzing the data with both pipelines (GstLAL and cWB) to produce a list of combined $p$-values $\overline{\mathcal{P}}$.  A simulated signal is considered to be detected by the search if $\overline{\mathcal{P}}$ is smaller than the $p$-value of the loudest event, $0.26$.  The sensitive $\langle VT\rangle$ is then given by
\begin{equation}
    \langle VT\rangle = T_{0} \int \,\dd z \,\dd\theta \frac{\dd V_\mathrm{c}}{\dd z}\frac{1}{1+z}s(\theta)f(z,\theta) \, ,
    \label{eq:VT}
\end{equation}
where $T_{0}$ is the total time covered by the injections (in the detector frame), $V_\mathrm{c}(z)$ is the comoving volume contained within a sphere out to redshift $z$~\cite{Hogg:1999}, $s(\theta)$ is the injected distribution of binary parameters $\theta$ (e.g., masses, spins, orientation angles, distance), and $0 \leq f(z,\theta) \leq 1$ is the selection function indicating the fraction of injections with redshift $z$ and parameters $\theta$ that are detected by our search.  We evaluate the integral~\eqref{eq:VT} using a Monte Carlo technique.

The injected waveforms are generated using a spin-aligned effective-one-body model~\cite{SEOBNRv2}, which is the waveform model used as a base for the reduced-order model~\cite{Purrer:2015tud} that the GstLAL search pipeline used for its template bank.   Precession and higher-order modes are possibly important for IMBHB detection~\cite{OShaughnessy:2012iol,Schmidt:2014iyl,Capano:2013raa,Varma:2014jxa,Harry:2016ijz,Bustillo:2015qty,Bustillo:2016gid}, particularly for sources with more extreme mass ratios; however, we neglect both effects due to current limitations in the waveform models.

Since the true population of IMBHBs, and thus the true function $s(\theta)$, is unknown, we focus on placing limits on twelve specific locations in the IMBHB parameter space.  We choose ten specific combinations of masses (see Table~\ref{tab1}).  For nine of these mass combinations, we consider only nonspinning black holes.  In the case 
$m_1 = m_2 = 100\,M_{\odot}$, we consider nonspinning black holes and two spinning cases.  
In both spinning cases, we choose dimensionless spins $\chi_{1,2}$ of magnitude $0.8$ which are aligned with each other.  In one case, the spins are also aligned with the orbital angular momentum of the system ($\chi_1 = \chi_2 = 0.8$); in the other, they are antialigned ($\chi_1 = \chi_2 = -0.8$).
Angular parameters (i.e., binary orientation and sky location) are chosen from a uniform distribution on a sphere.

The luminosity distances of the sources are chosen approximately uniformly in comoving volume out to a maximum redshift $z = 1$.\footnote{A flat cosmology with an incorrect value of $\Omega_\mathrm{m} = 0.3156$ (instead of $0.3065$) was used to generate the injection sets. We find that the error has no significant effect on our results, introducing an error of less than $1\%$.} The sources are distributed uniformly in the O1 observation time ($T_0 \simeq 130$ days), with a correction factor to account for time dilation.  In the detector frame, the injections are spaced by $100~\mathrm{s}$ on average.  The total number of injections in each set is $N_\mathrm{total}\simeq 112000$, with some slight variation between sets due to the random nature of assigning injection times.  Each set includes times during which the detectors were not taking coincident data; the procedure is insensitive to their inclusion in the total.  The total spacetime volume covered by the injection sets is  $\langle VT\rangle_\mathrm{total} \simeq 35~\mathrm{Gpc^3\,yr}$.  With these choices, expression~\eqref{eq:VT} for the sensitive $\langle VT\rangle$ reduces to the form
\begin{equation}
    \langle VT\rangle = \frac{N_\mathrm{below\ cutoff}}{N_\mathrm{total}}\langle VT\rangle_\mathrm{total},
\end{equation}
where $N_\mathrm{below\ cutoff}$ is the number 
of injections assigned a $p$-value lower than $0.26$.

The results are given in Table~\ref{tab1}. The table shows the $90\%$ confidence rate upper limit for each of
the twelve injection sets. Amplitude and phase errors arising from detector calibration~\cite{Abbott:2016jsd} have not been
included in the analysis; we expect uncertainty in $R_{90\%}$ to be $\approx 18\%$ because of the $\approx6\%$ uncertainty in the detectors' amplitude calibration~\cite{TheLIGOScientific:2016pea}. 
The tightest bound is placed on the merger of two $100\, M_\odot$ black holes whose spins are 
aligned with their orbital angular momentum: the rate of these mergers is constrained to be less than 
$0.93~\mathrm{Gpc^{-3}\, yr^{-1}}$.
Since IMBHB merger rates are commonly expressed in events per GC per Gyr, we convert our results into these units by assuming, for the sake of simplicity, a redshift-independent GC density of $3~\mathrm{GC\,Mpc^{-3}}$~\cite{PortegiesZwart:2000},\footnote{This density encompasses GCs with a range of masses and central concentrations; we make the further simplifying assumption that all GCs have the potential to form IMBHs with the masses we consider.} yielding a minimal $R_{90\%} \approx 0.3~\mathrm{GC^{-1}\, Gyr^{-1}}$.

We also report a sensitive distance,

\begin{equation}
  D_{\langle VT\rangle} = \left(\frac{3\langle VT\rangle}{4\pi T_\mathrm{a}}\right)^{1/3},
  \label{eq:D}
\end{equation}
where $T_\mathrm{a} < T_0$ is the total time analyzed by the search. The sensitive distance is analogous to the sense-monitor range~\cite{Allen:2012}, except that \eqref{eq:D} includes cosmological effects.
It is given in Table~\ref{tab1} for the combined
$\langle VT\rangle$, used to generate $R_{90\%}$, as well as for the GstLAL and cWB search algorithms individually.  
The searches are most sensitive to binaries with $m_1 = m_2 = 100\,M_\odot$ and aligned spins.  
Fig.~\ref{f:Rates} also reports $R_{90\%}$ and $ D_{\langle VT\rangle}$ for the combined search with lines of constant mass ratio $q=m_2/m_1$ and total mass $M=m_1+m_2$ to guide the eye.

\section{Astrophysical Implications}
\label{sec:astro}

There are currently few good candidates
for IMBHs, but if one extrapolates the observed relation between supermassive black holes and the masses of their host galaxies to lower-mass systems,
it is plausible to infer the existence of IMBHs~\cite{Ferrarese:2000se,Gebhardt:2000fk,2009ApJ...706..404G,2011MNRAS.412.2211G,2013ApJ...764..184M,1993IAUS..153..209K,Marconi:2003hj,2013ApJ...764..151G}.    While the formation channel of IMBHs is unknown, there are a small number of proposed scenarios: (i) the direct collapse of massive first-generation, low-metallicity Population III stars~\cite{Madau:2001sc,Schneider:2001bu,Ryu:2016dou,2014MNRAS.442.2963K}, (ii)
runaway mergers of massive main sequence stars in dense stellar clusters~\cite{Miller:2001ez,PortegiesZwart:2004ggg,PortegiesZwart:2002iks,AtakanGurkan:2003hm,Mapelli:2016vca}, (iii) the accretion of residual gas onto stellar-mass black holes~\cite{2013MNRAS.433.1958L} and (iv) chemically homogeneous evolution~\cite{Marchant:2016wow}.\footnote{Since IMBHs are potentially formed via different channels than stellar-mass black holes, we do not attempt to extrapolate the BBH mass distribution to IMBHBs. The O1 BBH merger rate and mass distribution reported in \cite{Abbott:2016nhf,TheLIGOScientific:2016pea} were calculated assuming that the total mass is less than $100\,M_{\odot}$.}  

It has been suggested that the most likely locations to find IMBHs are at the centers of 
GCs
~\cite{Godet:2009hn,2017MNRAS.464.3090A,Lanzoni:2007ms,2010AIPC.1240..245J,Noyola:2008kt,2010ApJ...710.1032A,Pasham:2015tca,Brightman:2016pax,Cseh:2014qsa,Mezcua:2015pra,Lutzgendorf:2013csa,2016arXiv161206794K,2017arXiv170202149K}.      It follows that these are also the most likely places to find IMBHBs. Again, while the formation mechanisms are unknown, it is postulated that an IMBHB can be formed in a GC with a fraction of binary stars higher than $\approx10\%$~\cite{Gurkan:2005xz} or as a result of a merger of two clusters, each of which contains an IMBH~\cite{AmaroSeoane:2006py,Fregeau:2006yz}.  While no direct observational evidence of IMBHBs exists, this hypothesis is supported by recent simulations of dense stellar systems~\cite{2015MNRAS.454.3150G}.  Measurements of an 
IMBHB's components would allow us to not only constrain IMBH formation channels, but also make statements on the link between IMBHs and
both ultra-~\cite{2013MNRAS.436.3128M} and hyper-luminous~\cite{Gao:2003sr,2009Natur.460...73F,2012MNRAS.423.1154S} X-ray systems. 

As stated in Table~\ref{tab1}, the minimal $R_{90\%}$ is found to be $\approx0.3~\mathrm{GC^{-1}\,Gyr}^{-1}$.  The improvement in detector sensitivity since the S6 run means that this result is nearly two orders of magnitude lower than the lowest upper limit set using previous LIGO--Virgo data~\cite{Aasi:2014bqj, Aasi:2014iwa}. This number is within a factor of a few of
$0.1~\mathrm{GC^{-1}\,Gyr}^{-1}$, the IMBHB merger rate corresponding to one event occurring in each GC within the lifetime of the cluster (assumed equal to $10~\mathrm{Gyr}$), although it only refers to a single point in the IMBHB mass--spin parameter space and not to the full physical distribution of IMBHBs. The bounds are compatible with rate predictions coming from astrophysical models
of IMBHB formation~\cite{Fregeau:2006yz,2016MmSAI..87..555G,Rodriguez:2016kxx}. To make a full comparison of the upper limits with predictions, or with the BBH merger rate ($9$--$240~\mathrm{Gpc^{-3}\,{yr}^{-1}}$~\cite{Abbott:2016nhf,TheLIGOScientific:2016pea}), it would be necessary to assume a mass, spin and redshift distribution for IMBHB mergers; this distribution is currently uncertain, so we defer a comparison to future studies.

Further improvements to the detector sensitivity in the next observing runs will allow us to increasingly improve the IMBHB merger-rate estimation and provide relevant constraints on the merger rate in the local Universe.   A single GW detection of an IMBHB merger could provide the first conclusive proof of the existence of IMBHs in the Universe~\cite{Veitch:2015ela,Graff:2015bba,Haster:2015cnn}.  Multiple detections, where astrophysically important parameters, such as mass and spin,
are measured, would allow us to make statements not only on the formation and evolutionary channels of IMBHs, but also on their link with other
observed phenomena.

\section{Conclusion}
\label{sec:conclusion}
This paper describes a search for intermediate mass black hole binaries during the first observing run of the Advanced LIGO detectors. Due to improvement in detector sensitivity, this run had an increase in search horizon of a factor of $\approx6$ compared to the previous science run.  The search uses the combined information
from a modeled matched-filter pipeline and an unmodeled transient burst pipeline.  
While no IMBHBs were found, $90\%$ confidence limits were placed on the merger rates of IMBHBs in the local universe. The minimum 
merger rate of $\approx0.3~\mathrm{GC^{-1}\,Gyr}^{-1}$ constitutes an improvement of almost two orders of magnitude over
the previous search results.  The results presented here are based on nonprecessing and, in most cases, nonspinning waveforms, that also
omit higher modes.  It is believed that these higher-order physical effects may be important for IMBHBs, but they should be less important for the near equal-mass systems where we can set best upper limits. We plan to include these effects in future analyses.
It is also believed that continued improvements in the detector performance during future observing runs~\cite{Aasi:2013wya} will allow us
to further tighten these bounds and may lead to the first detections of IMBHs.

\section*{Acknowledgments} 

The authors gratefully acknowledge the support of the United States
National Science Foundation (NSF) for the construction and operation of the
LIGO Laboratory and Advanced LIGO as well as the Science and Technology Facilities Council (STFC) of the
United Kingdom, the Max-Planck-Society (MPS), and the State of
Niedersachsen/Germany for support of the construction of Advanced LIGO 
and construction and operation of the GEO600 detector. 
Additional support for Advanced LIGO was provided by the Australian Research Council.
The authors gratefully acknowledge the Italian Istituto Nazionale di Fisica Nucleare (INFN),  
the French Centre National de la Recherche Scientifique (CNRS) and
the Foundation for Fundamental Research on Matter supported by the Netherlands Organisation for Scientific Research, 
for the construction and operation of the Virgo detector
and the creation and support  of the EGO consortium. 
The authors also gratefully acknowledge research support from these agencies as well as by 
the Council of Scientific and Industrial Research of India, 
Department of Science and Technology, India,
Science \& Engineering Research Board (SERB), India,
Ministry of Human Resource Development, India,
the Spanish Ministerio de Econom\'ia y Competitividad,
the  Vicepresid\`encia i Conselleria d'Innovaci\'o, Recerca i Turisme and the Conselleria d'Educaci\'o i Universitat del Govern de les Illes Balears,
the National Science Centre of Poland,
the European Commission,
the Royal Society, 
the Scottish Funding Council, 
the Scottish Universities Physics Alliance, 
the Hungarian Scientific Research Fund (OTKA),
the Lyon Institute of Origins (LIO),
the National Research Foundation of Korea,
Industry Canada and the Province of Ontario through the Ministry of Economic Development and Innovation, 
the Natural Science and Engineering Research Council Canada,
Canadian Institute for Advanced Research,
the Brazilian Ministry of Science, Technology, and Innovation,
International Center for Theoretical Physics South American Institute for Fundamental Research (ICTP-SAIFR), 
Russian Foundation for Basic Research,
the Leverhulme Trust, 
the Research Corporation, 
Ministry of Science and Technology (MOST), Taiwan
and
the Kavli Foundation.
The authors gratefully acknowledge the support of the NSF, STFC, MPS, INFN, CNRS and the
State of Niedersachsen/Germany for provision of computational resources.
This is LIGO document LIGO-P1600273.

\bibliography{references}

\iftoggle{endauthorlist}{
  %
  %
  \let\author\myauthor
  \let\affiliation\myaffiliation
  \let\maketitle\mymaketitle
  \title{Authors}
  \pacs{}

  \author{%
B.~P.~Abbott,$^{1}$  
R.~Abbott,$^{1}$  
T.~D.~Abbott,$^{2}$  
F.~Acernese,$^{3,4}$ 
K.~Ackley,$^{5}$  
C.~Adams,$^{6}$  
T.~Adams,$^{7}$ 
P.~Addesso,$^{8}$  
R.~X.~Adhikari,$^{1}$  
V.~B.~Adya,$^{9}$
% ...
\\
\medskip
(LIGO Scientific Collaboration and Virgo Collaboration) 
\\
\medskip
{{}$^{*}$Deceased, March 2016. }%
{{}$^{\sharp}$Deceased, March 2017. }%
{${}^{\dag}$Deceased, February 2017. }%
{${}^{\ddag}$Deceased, December 2016. }%
}\noaffiliation
\affiliation {LIGO, California Institute of Technology, Pasadena, CA 91125, USA }
\affiliation {Louisiana State University, Baton Rouge, LA 70803, USA }
\affiliation {Universit\`a di Salerno, Fisciano, I-84084 Salerno, Italy }
\affiliation {INFN, Sezione di Napoli, Complesso Universitario di Monte S.Angelo, I-80126 Napoli, Italy }
\affiliation {University of Florida, Gainesville, FL 32611, USA }
\affiliation {LIGO Livingston Observatory, Livingston, LA 70754, USA }
\affiliation {Laboratoire d'Annecy-le-Vieux de Physique des Particules (LAPP), Universit\'e Savoie Mont Blanc, CNRS/IN2P3, F-74941 Annecy, France }
\affiliation {University of Sannio at Benevento, I-82100 Benevento, Italy and INFN, Sezione di Napoli, I-80100 Napoli, Italy }
\affiliation {Albert-Einstein-Institut, Max-Planck-Institut f\"ur Gravi\-ta\-tions\-physik, D-30167 Hannover, Germany }
\affiliation {The University of Mississippi, University, MS 38677, USA }
\affiliation {NCSA, University of Illinois at Urbana-Champaign, Urbana, IL 61801, USA }
\affiliation {Nikhef, Science Park, 1098 XG Amsterdam, The Netherlands }
\affiliation {LIGO, Massachusetts Institute of Technology, Cambridge, MA 02139, USA }
\affiliation {Instituto Nacional de Pesquisas Espaciais, 12227-010 S\~{a}o Jos\'{e} dos Campos, S\~{a}o Paulo, Brazil }
\affiliation {Gran Sasso Science Institute (GSSI), I-67100 L'Aquila, Italy }
\affiliation {INFN, Sezione di Roma Tor Vergata, I-00133 Roma, Italy }
\affiliation {Inter-University Centre for Astronomy and Astrophysics, Pune 411007, India }
\affiliation {University of Wisconsin-Milwaukee, Milwaukee, WI 53201, USA }
\affiliation {Leibniz Universit\"at Hannover, D-30167 Hannover, Germany }
\affiliation {Universit\`a di Pisa, I-56127 Pisa, Italy }
\affiliation {INFN, Sezione di Pisa, I-56127 Pisa, Italy }
\affiliation {SUPA, University of Glasgow, Glasgow G12 8QQ, United Kingdom }
\affiliation {Australian National University, Canberra, Australian Capital Territory 0200, Australia }
\affiliation {Laboratoire des Mat\'eriaux Avanc\'es (LMA), CNRS/IN2P3, F-69622 Villeurbanne, France }
\affiliation {LAL, Univ. Paris-Sud, CNRS/IN2P3, Universit\'e Paris-Saclay, F-91898 Orsay, France }
\affiliation {California State University Fullerton, Fullerton, CA 92831, USA }
\affiliation {European Gravitational Observatory (EGO), I-56021 Cascina, Pisa, Italy }
\affiliation {Chennai Mathematical Institute, Chennai 603103, India }
\affiliation {Universit\`a di Roma Tor Vergata, I-00133 Roma, Italy }
\affiliation {Universit\"at Hamburg, D-22761 Hamburg, Germany }
\affiliation {INFN, Sezione di Roma, I-00185 Roma, Italy }
\affiliation {Embry-Riddle Aeronautical University, Prescott, AZ 86301, USA }
\affiliation {Albert-Einstein-Institut, Max-Planck-Institut f\"ur Gravitations\-physik, D-14476 Potsdam-Golm, Germany }
\affiliation {APC, AstroParticule et Cosmologie, Universit\'e Paris Diderot, CNRS/IN2P3, CEA/Irfu, Observatoire de Paris, Sorbonne Paris Cit\'e, F-75205 Paris Cedex 13, France }
\affiliation {Korea Institute of Science and Technology Information, Daejeon 34141, Korea }
\affiliation {West Virginia University, Morgantown, WV 26506, USA }
\affiliation {Center for Gravitational Waves and Cosmology, West Virginia University, Morgantown, WV 26505, USA }
\affiliation {Universit\`a di Perugia, I-06123 Perugia, Italy }
\affiliation {INFN, Sezione di Perugia, I-06123 Perugia, Italy }
\affiliation {Syracuse University, Syracuse, NY 13244, USA }
\affiliation {University of Minnesota, Minneapolis, MN 55455, USA }
\affiliation {LIGO Hanford Observatory, Richland, WA 99352, USA }
\affiliation {Wigner RCP, RMKI, H-1121 Budapest, Konkoly Thege Mikl\'os \'ut 29-33, Hungary }
\affiliation {Columbia University, New York, NY 10027, USA }
\affiliation {Stanford University, Stanford, CA 94305, USA }
\affiliation {Universit\`a di Camerino, Dipartimento di Fisica, I-62032 Camerino, Italy }
\affiliation {Universit\`a di Padova, Dipartimento di Fisica e Astronomia, I-35131 Padova, Italy }
\affiliation {INFN, Sezione di Padova, I-35131 Padova, Italy }
\affiliation {MTA E\"otv\"os University, ``Lendulet'' Astrophysics Research Group, Budapest 1117, Hungary }
\affiliation {Nicolaus Copernicus Astronomical Center, Polish Academy of Sciences, 00-716, Warsaw, Poland }
\affiliation {University of Birmingham, Birmingham B15 2TT, United Kingdom }
\affiliation {Universit\`a degli Studi di Genova, I-16146 Genova, Italy }
\affiliation {INFN, Sezione di Genova, I-16146 Genova, Italy }
\affiliation {RRCAT, Indore MP 452013, India }
\affiliation {Faculty of Physics, Lomonosov Moscow State University, Moscow 119991, Russia }
\affiliation {SUPA, University of the West of Scotland, Paisley PA1 2BE, United Kingdom }
\affiliation {Caltech CaRT, Pasadena, CA 91125, USA }
\affiliation {University of Western Australia, Crawley, Western Australia 6009, Australia }
\affiliation {Department of Astrophysics/IMAPP, Radboud University Nijmegen, P.O. Box 9010, 6500 GL Nijmegen, The Netherlands }
\affiliation {Artemis, Universit\'e C\^ote d'Azur, Observatoire C\^ote d'Azur, CNRS, CS 34229, F-06304 Nice Cedex 4, France }
\affiliation {Institut de Physique de Rennes, CNRS, Universit\'e de Rennes 1, F-35042 Rennes, France }
\affiliation {Washington State University, Pullman, WA 99164, USA }
\affiliation {Universit\`a degli Studi di Urbino 'Carlo Bo', I-61029 Urbino, Italy }
\affiliation {INFN, Sezione di Firenze, I-50019 Sesto Fiorentino, Firenze, Italy }
\affiliation {University of Oregon, Eugene, OR 97403, USA }
\affiliation {Laboratoire Kastler Brossel, UPMC-Sorbonne Universit\'es, CNRS, ENS-PSL Research University, Coll\`ege de France, F-75005 Paris, France }
\affiliation {Carleton College, Northfield, MN 55057, USA }
\affiliation {Astronomical Observatory Warsaw University, 00-478 Warsaw, Poland }
\affiliation {VU University Amsterdam, 1081 HV Amsterdam, The Netherlands }
\affiliation {University of Maryland, College Park, MD 20742, USA }
\affiliation {Center for Relativistic Astrophysics and School of Physics, Georgia Institute of Technology, Atlanta, GA 30332, USA }
\affiliation {Universit\'e Claude Bernard Lyon 1, F-69622 Villeurbanne, France }
\affiliation {Universit\`a di Napoli 'Federico II', Complesso Universitario di Monte S.Angelo, I-80126 Napoli, Italy }
\affiliation {NASA Goddard Space Flight Center, Greenbelt, MD 20771, USA }
\affiliation {RESCEU, University of Tokyo, Tokyo, 113-0033, Japan. }
\affiliation {University of Adelaide, Adelaide, South Australia 5005, Australia }
\affiliation {Tsinghua University, Beijing 100084, China }
\affiliation {Texas Tech University, Lubbock, TX 79409, USA }
\affiliation {Kenyon College, Gambier, OH 43022, USA }
\affiliation {The Pennsylvania State University, University Park, PA 16802, USA }
\affiliation {National Tsing Hua University, Hsinchu City, 30013 Taiwan, Republic of China }
\affiliation {Charles Sturt University, Wagga Wagga, New South Wales 2678, Australia }
\affiliation {University of Chicago, Chicago, IL 60637, USA }
\affiliation {Pusan National University, Busan 46241, Korea }
\affiliation {University of Cambridge, Cambridge CB2 1TN, United Kingdom }
\affiliation {The Chinese University of Hong Kong, Shatin, NT, Hong Kong }
\affiliation {INAF, Osservatorio Astronomico di Padova, Vicolo dell'Osservatorio 5, I-35122 Padova, Italy }
\affiliation {INFN, Trento Institute for Fundamental Physics and Applications, I-38123 Povo, Trento, Italy }
\affiliation {Universit\`a di Roma 'La Sapienza', I-00185 Roma, Italy }
\affiliation {Universit\'e Libre de Bruxelles, Brussels 1050, Belgium }
\affiliation {Sonoma State University, Rohnert Park, CA 94928, USA }
\affiliation {Montana State University, Bozeman, MT 59717, USA }
\affiliation {Center for Interdisciplinary Exploration \& Research in Astrophysics (CIERA), Northwestern University, Evanston, IL 60208, USA }
\affiliation {Cardiff University, Cardiff CF24 3AA, United Kingdom }
\affiliation {Universitat de les Illes Balears, IAC3---IEEC, E-07122 Palma de Mallorca, Spain }
\affiliation {The University of Texas Rio Grande Valley, Brownsville, TX 78520, USA }
\affiliation {Bellevue College, Bellevue, WA 98007, USA }
\affiliation {Institute for Plasma Research, Bhat, Gandhinagar 382428, India }
\affiliation {The University of Sheffield, Sheffield S10 2TN, United Kingdom }
\affiliation {University of Southern California Information Sciences Institute, Marina Del Rey, CA 90292, USA }
\affiliation {California State University, Los Angeles, 5151 State University Dr, Los Angeles, CA 90032, USA }
\affiliation {Universit\`a di Trento, Dipartimento di Fisica, I-38123 Povo, Trento, Italy }
\affiliation {Montclair State University, Montclair, NJ 07043, USA }
\affiliation {National Astronomical Observatory of Japan, 2-21-1 Osawa, Mitaka, Tokyo 181-8588, Japan }
\affiliation {Canadian Institute for Theoretical Astrophysics, University of Toronto, Toronto, Ontario M5S 3H8, Canada }
\affiliation {Whitman College, 345 Boyer Avenue, Walla Walla, WA 99362 USA }
\affiliation {School of Mathematics, University of Edinburgh, Edinburgh EH9 3FD, United Kingdom }
\affiliation {University and Institute of Advanced Research, Gandhinagar Gujarat 382007, India }
\affiliation {IISER-TVM, CET Campus, Trivandrum Kerala 695016, India }
\affiliation {University of Szeged, D\'om t\'er 9, Szeged 6720, Hungary }
\affiliation {International Centre for Theoretical Sciences, Tata Institute of Fundamental Research, Bengaluru 560089, India }
\affiliation {Tata Institute of Fundamental Research, Mumbai 400005, India }
\affiliation {INAF, Osservatorio Astronomico di Capodimonte, I-80131, Napoli, Italy }
\affiliation {University of Michigan, Ann Arbor, MI 48109, USA }
\affiliation {American University, Washington, D.C. 20016, USA }
\affiliation {Rochester Institute of Technology, Rochester, NY 14623, USA }
\affiliation {University of Bia{\l }ystok, 15-424 Bia{\l }ystok, Poland }
\affiliation {SUPA, University of Strathclyde, Glasgow G1 1XQ, United Kingdom }
\affiliation {University of Southampton, Southampton SO17 1BJ, United Kingdom }
\affiliation {University of Washington Bothell, 18115 Campus Way NE, Bothell, WA 98011, USA }
\affiliation {Institute of Applied Physics, Nizhny Novgorod, 603950, Russia }
\affiliation {Seoul National University, Seoul 08826, Korea }
\affiliation {Inje University Gimhae, South Gyeongsang 50834, Korea }
\affiliation {National Institute for Mathematical Sciences, Daejeon 34047, Korea }
\affiliation {NCBJ, 05-400 \'Swierk-Otwock, Poland }
\affiliation {Institute of Mathematics, Polish Academy of Sciences, 00656 Warsaw, Poland }
\affiliation {The School of Physics \& Astronomy, Monash University, Clayton 3800, Victoria, Australia }
\affiliation {Hanyang University, Seoul 04763, Korea }
\affiliation {NASA Marshall Space Flight Center, Huntsville, AL 35811, USA }
\affiliation {ESPCI, CNRS, F-75005 Paris, France }
\affiliation {Southern University and A\&M College, Baton Rouge, LA 70813, USA }
\affiliation {The University of Melbourne, Parkville, Victoria 3010, Australia }
\affiliation {College of William and Mary, Williamsburg, VA 23187, USA }
\affiliation {Indian Institute of Technology Madras, Chennai 600036, India }
\affiliation {IISER-Kolkata, Mohanpur, West Bengal 741252, India }
\affiliation {Scuola Normale Superiore, Piazza dei Cavalieri 7, I-56126 Pisa, Italy }
\affiliation {Universit\'e de Lyon, F-69361 Lyon, France }
\affiliation {Hobart and William Smith Colleges, Geneva, NY 14456, USA }
\affiliation {Janusz Gil Institute of Astronomy, University of Zielona G\'ora, 65-265 Zielona G\'ora, Poland }
\affiliation {University of Washington, Seattle, WA 98195, USA }
\affiliation {King's College London, University of London, London WC2R 2LS, United Kingdom }
\affiliation {Indian Institute of Technology, Gandhinagar Ahmedabad Gujarat 382424, India }
\affiliation {International Institute of Physics, Universidade Federal do Rio Grande do Norte, Natal RN 59078-970, Brazil }
\affiliation {Andrews University, Berrien Springs, MI 49104, USA }
\affiliation {Universit\`a di Siena, I-53100 Siena, Italy }
\affiliation {Trinity University, San Antonio, TX 78212, USA }
\affiliation {Abilene Christian University, Abilene, TX 79699, USA }

  \newpage
  \maketitle
}

\end{document}